\documentclass[%
 aip,
 amsmath,amssymb,
 preprint,%
]{revtex4-1}

\usepackage{graphicx}
\usepackage{dcolumn}
\usepackage{bm}

\usepackage[utf8]{inputenc}
\usepackage[T1]{fontenc}
\usepackage{mathptmx}
\usepackage{siunitx}
\usepackage{braket}
\begin{document}


\title[Resonant Tunneling Diodes Strongly Coupled to the Cavity Field]{Resonant Tunneling Diodes Strongly Coupled to the Cavity Field}

\author{B.~Limbacher}
 \email{benedikt.limbacher@tuwien.ac.at}
\author{M.~A.~Kainz}
\author{S.~Schoenhuber}
\author{M.~Wenclawiak}
\author{C.~Derntl}
\affiliation{ 
Photonics Institute, TU Wien, Gusshausstrasse 27-29, Vienna, Austria
}
\affiliation{%
Center for Micro- and Nanostructures, TU Wien, Gusshausstrasse 25-25a, Vienna, Austria
}%

\author{A.~M.~Andrews}
\affiliation{%
Center for Micro- and Nanostructures, TU Wien, Gusshausstrasse 25-25a, Vienna, Austria
}%
\affiliation{%
Institute of Solid State Electronics, TU Wien, Gusshausstrasse 25-25a, Vienna, Austria
}%

\author{H.~Detz}
\affiliation{%
Center for Micro- and Nanostructures, TU Wien, Gusshausstrasse 25-25a, Vienna, Austria
}%
\affiliation{%
Central European Institute of Technology, Purkynova 123, Brno, Czech Republic}%

\author{G.~Strasser}
\affiliation{%
Center for Micro- and Nanostructures, TU Wien, Gusshausstrasse 25-25a, Vienna, Austria
}%
\affiliation{%
Institute of Solid State Electronics, TU Wien, Gusshausstrasse 25-25a, Vienna, Austria
}%

\author{A.~Schwaighofer}
\author{B.~Lendl}
\affiliation{Institute of Chemical Technologies and Analytics, TU Wien, Getreidemarkt 9/164, Vienna, Austria}

\author{J.~Darmo}
\author{K.~Unterrainer}
\affiliation{ 
Photonics Institute, TU Wien, Gusshausstrasse 27-29, Vienna, Austria
}
\affiliation{%
Center for Micro- and Nanostructures, TU Wien, Gusshausstrasse 25-25a, Vienna, Austria
}%
\date{\today}

\begin{abstract}
We demonstrate Resonant Tunneling Diodes, embedded in double metal cavities, strongly coupled to the cavity field, while maintaining their electronic properties. We measure the polariton dispersion and find a relative vacuum Rabi splitting of 16\%, which explicitly qualifies for the strong-coupling regime. Additionally we show that electronic transport has a significant influence on the polaritons by modulating the coupling strength. The merge between electronic transport and polaritonic physics in our devices opens up a new aspect of cavity quantum electro-dynamics and integrated photonics.
\end{abstract}

\maketitle

The emergence of semiconductor heterostructures has enabled countless electrical and optical applications such as resonant tunneling diodes \cite{tsu_tunneling_1973}, quantum cascade lasers \cite{faist_quantum_1994}, quantum well infrared photodetectors \cite{levine_quantum-well_1993, noda_all-optical_1990} and high speed optical modulators \cite{wood_highspeed_1984}. This led to technological access of frequency ranges beyond the reach of material bandgaps. A major advantage of these systems is that the transitions between the electron energy levels (intersubband transitions), which originate from the electronic confinement in one direction, can be engineered. By embedding heterostructures in an optical cavity, which is in resonance with the intersubband transition,  strong light-matter interaction occurs \cite{ciuti_quantum_2005}. Interesting physical phenomena, such as the vacuum Rabi splitting \cite{liu_rabi_1997, khitrova_vacuum_2006} can be observed in this regime. The occurrence of the vacuum Rabi splitting has been reported for various frequency ranges, covering the microwave, THz  \cite{dietze_terahertz_2011} and mid-infrared \cite{dini_microcavity_2003} regions. Even though modulation of the reflection spectrum of semiconductor heterostructures \cite{sarma_metasurface_2018, anappara_tunnel-assisted_2006}, electroluminescence from intersubband polaritons \cite{sapienza_electrically_2008} and modified absorption spectra in strongly coupled quantum well infrared photodetectors \cite{vigneron_quantum_2019} have been shown, electronic transport in strongly coupled systems has mostly been neglected in scientific investigations. Due to the large intersubband matrix element, the Rabi splitting in modulation doped semiconductor nanostructures is very large and even the quantum Hall effect regime can be modified \cite{paravicini-bagliani_magneto-transport_2019}.

Until today, resonant tunneling diodes, which are ideal systems for investigating resonant electronic transport, have been operating in the weak light-matter coupling regime. The carrier densities in resonant tunneling diodes can reach the same magnitude as in modulation doped structures and can be controlled very fast (less than picoseconds) by the current due to large tunneling rates \cite{feiginov_operation_2014}. Here, we demonstrate room temperature operation of triple barrier resonant tunneling diodes \cite{nakagawa_sharp_1987}, strongly coupled to the field of a cavity at mid-infrared frequencies. The cavity is resonant to the lowest intersubband transition of the quantum wells. We show, that electronic transport leads to a significant modulation of the coupling strength due to a modulation of the electron injection and extraction rates of the system. This has to be considered when designing transport enabled polaritonic devices. Our results pave the way for gaining electrical control of intersubband polaritons. In addition, our results create new perspectives for investigating electronic transport in strongly coupled systems, which is crucial for novel polaritonic applications such as intersubband polariton lasers \cite{colombelli_perspectives_2015}.

The dominant parameter for strongly coupled systems is the coupling strength $\Omega_{R}$, which is also know as the Rabi-frequency. Only for coupling strengths on the order of a few percent of the transition energy or higher, strong-coupling effects can be observed, as the coupling strength has to outweigh the system losses. 
The coupling strength is essentially given by the transition dipole element $\vec d_{12}$ of the intersubband transition and the electric field $\vec E_{cav}$ in the cavity \cite{khitrova_vacuum_2006}.
By using information about the geometry the coupling strength can be written as
\begin{equation}
\label{eq:omega1}
\hbar \Omega_{R} = \vec d_{12}\cdot \vec E_{cav} \propto \sqrt{\frac{N_{2DEG} n_{QW}f_{12}}{V}},
\end{equation}
where the product of the 2D electron gas carrier density $N_{2DEG}$ and the number of quantum wells $n_{QW}$ defines the number of charge carriers available for coupling. The $f_{12}$ is the oscillator strength of the lowest order intersubband transition. The cavity field scales inversely proportional with the mode volume $V$. Hence, small cavities and materials with many charge carriers are favorable for boosting the coupling strength. Resonant tunneling diodes are usually fabricated as small devices to reduce the time constant and current instabilities. Thus, the scaling for high performance devices favors the occurrence of strong coupling. The modulation of the number of charge carriers, which couple to the respective cavity mode, leads to a modulation of the coupling strength, which can be observed by a change of the minimal energy separation between the lower and upper polariton absorption peaks.
%
%
%
The heterostructure used for the experiments is a GaAs/AlGaAs based triple barrier resonant tunneling diode, grown by molecular beam epitaxy. This design is better suited for strong-coupling experiments than the more common double barrier resonant tunneling diode designs \cite{chang_resonant_1974}, as the two quantum wells effectively enhance the coupling strength. The devices show an asymmetric electrical behavior caused by an undoped spacer region attached to one side of the quantum wells \cite{nakagawa_sharp_1987}.

\begin{figure}
\centering\includegraphics[width=7cm]{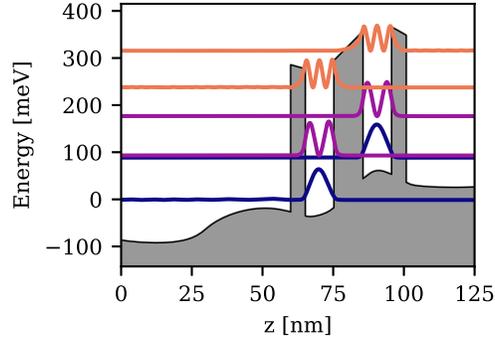}
\caption{Simulated conduction band of the triple barrier resonant tunneling diode with an electric field applied, so that the ground-state of the first well and the first-excited state of the second well align. The blue curves are the ground-states of the quantum wells, the violet curves are the first-excited states. The second-excited states are depicted in orange. The grey shaded area is the heterostructure band-gap. The transition energy between the ground and excited states is calculated to be 95 \si{\milli \electronvolt} and the transition matrix element is estimated to be 2.5 \si{\nano \metre}.}
\label{fig:sp}
\end{figure}

Our design features 70 \si{\nano \metre} wide, symmetrically n-doped contact layers with a sheet density of $7\times 10^{12}$ \si{\centi \metre ^{-2}}, leading to a homogeneous electric field distribution in the cavity. Additionally, the wells, with a width of 10.2 \si{\nano \metre} each, are n-doped with a sheet density of $N_{2DEG} = 1.2 \times 10^{12}$  \si{\centi \metre ^{-2}}, which increases the number of charge carriers available for coupling. The wells are separated from each other with a 10.2 \si{\nano \metre} wide $\mathrm{Al_{0.4}Ga_{0.6}As}$ barrier and from the contact layers with 5.1 \si{\nano \metre} wide $\mathrm{Al_{0.4}Ga_{0.6}As}$ barriers. Fig. \ref{fig:sp} shows the simulated conduction band of such a device with an electric field applied, so that the ground state of the first well and the first excited state of the second well are aligned. The intersubband transition energy between the ground and first excited states are calculated to be 95 \si{\milli \electronvolt}. The intersubband transition matrix element $f_{12}=\braket{0|z|1}$, where $\ket{0}$ is the ground state, $\ket{1}$ is the first excited state and $z$ is the position operator, is calculated to be 2.5 \si{\nano \metre}. Our simulations show that the perturbation of the transition matrix element for the entire operational range is non-significant. Note that the electron transit time $\tau_d = L^2/(\mu_e V_B)$, where $L$ is the device thickness, $\mu_e$ is the electron mobility in GaAs and $V_B$ is the applied voltage, is in the range of 100 \si{\femto \second} for almost the entire operational range of the device, which is considerably faster than the Rabi frequency ($\approx 2.4$ \si{\tera \hertz}). This can be identified as one of the main advantages of resonant tunneling diodes compared to multi quantum well structures, as they intrinsically allow to reach the nonadiabatic regime. The heterostructure is designed to have a high resistance and hence low current densities in order to avoid lateral electric field inhomogeneities in larger devices such as arrays. The high resistance has a negative impact on the maximum modulation frequency, due to an increase of the RC time constant. This can be counteracted by employing H-shaped resonators, which have a considerably lower electrostatic capacity than patch resonators.
%
%

\begin{figure}
\centering\includegraphics[width=8.5cm]{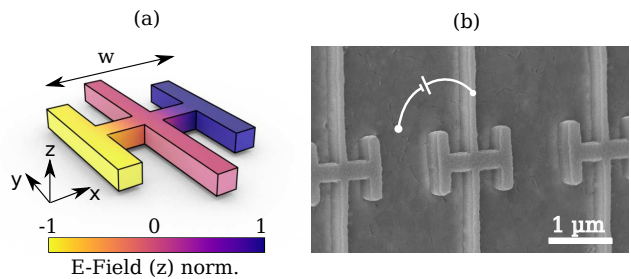}
\caption{Geometrically tunable resonator arrays allow the investigation of intersubband polaritons. (a) Geometry of a single H-shaped resonator with the simulated vertically polarized component of the electric field depicted in colors. The distance between the two outer bars $w$ (ranging from 1.14 \si{\micro \metre} to 1.72 \si{\micro \metre}) is used for tuning the resonance frequency. The outer bars themselves are 1 \si{\micro \metre} long and 250 \si{\nano \metre} wide. The contact line, connecting the individual resonators, is designed to be 150 \si{\nano \metre} wide. (b) Scanning electron microscope image of a resonator array. The single resonators are electrically connected via connection lines perpendicular to the fundamental cavity mode. The source symbol indicates that the current is flowing through the structure vertically from the top- to the bottom-contact.}
\label{fig:sem}
\end{figure}

Arrays of triple barrier resonant tunneling diodes, embedded in double metal H-shaped resonators, act as an active metamaterial surface (metasurface). The use of metasurfaces drastically increases the interaction area of the devices, compared to a single resonator. The top- and bottom metal layers additionally serve as electrical contacts for the resonant tunneling diodes. The regularly arranged microresonators, each of which consists of two identical, vertical 1 \si{\micro \metre} long bars with a width of 250 \si{\nano \metre} and a horizontal bar of the same width, with variable length, ranging from 1.14 \si{\micro \metre} to 1.72 \si{\micro \metre}, are electrically connected to each other via 150 \si{\nano \metre} wide connection lines perpendicular to the fundamental cavity mode dipole. The resonator arrays are fabricated by using electron beam lithography. A silicon nitride layer serves as electrical insulation between the ground plane and the contact lines, which connect the external bond pads and the resonator arrays. Windows structured in the silicon nitride layer allow unperturbed optical access to the arrays. Due to the intersubband transition selection rules, only the component of the electric field polarized in growth direction is absorbed. A simulation of the electric field, polarized in this direction, is depicted in Fig. \ref{fig:sem} (a). A scanning electron microscope picture of the processed sample is seen in Fig. \ref{fig:sem} (b). Eleven arrays, each of which has a different resonator width $w$, for different resonance frequencies were fabricated with an area of 200 \si{\micro \metre} by 200 \si{\micro \metre} each. The number of resonators per array ranges from 2000 to 10000. The resonance frequency is tuned by the distance $w$ of the bars of the H-shaped resonator (compare Fig. \ref{fig:sem} (a)). The array allows access to optical characterization techniques with a high signal to noise ratio, considering that the free-space wavelength is one order of magnitude larger than the size of a single resonator.

\begin{figure}
\centering\includegraphics[width=8cm]{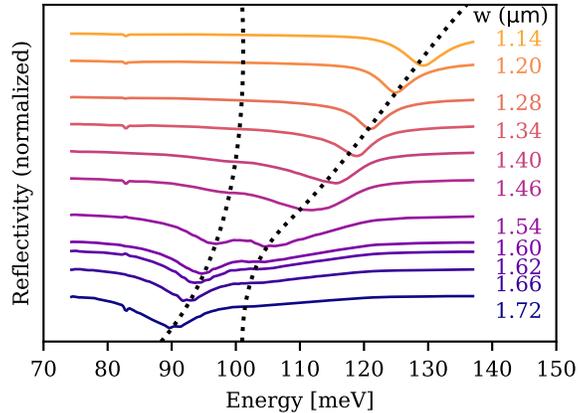}
\caption{Normalized reflection spectra of different resonator arrays with respective resonator sizes. The vacuum Rabi splitting near 96 \si{\milli \electronvolt} can be clearly identified. The dotted lines are guides to the eye, indicating the position of the polariton absorption peaks.}
\label{fig:coupling}
\end{figure}

A TM-polarized multipass measurement \cite{levine_quantum-well_1993} of the sample in a waveguide geometry allows a spectral characterization of the intersubband transitions. The sample substrate of 650 \si{\micro \metre} thickness has facets polished at a 45 degree angle and a length of 5 \si{\milli \metre}. This results in a total of 8 beam passes through the heterostructure, while at the same time the electric field fulfills the intersubband polarization selection rule. Our triple barrier resonant tunneling diodes show an intersubband transition energy of 96 \si{\milli \electronvolt} with a full width at half maximum linewidth of 6 \si{\milli \electronvolt} and a reflection modulation depth of 17\%. The voltage-current characteristics of etched double metal mesa devices, which are mounted in a continuous flow helium cryostat, show distinct resonant tunneling peaks up to operating temperatures of 150 \si{\kelvin}. The current densities at the lowest resonance are ranging from 3 to 9 \si{\ampere\per\square\centi\metre}, depending on the operating temperature.
Multiple arrays of resonators with varying resonance frequencies allow the measurement of the intersubband polariton dispersion as depicted in Fig. \ref{fig:coupling}. The data are recorded using a Bruker Hyperion 3000 Cassegrain type microscope attached to a Bruker Tensor 37 Fourier-Transform  Infrared (FTIR) spectrometer. A liquid nitrogen cooled mercury cadmium telluride detector is used for detection. We measure the reflection spectra at normal incidence using a beamsplitter, mounted within the FTIR-microscope. The spot size can be set from the full array size down to the diffraction limit. By measuring different subsections of the arrays we find that the spatial inhomogeneity of our device arrays is negligible.
The vacuum Rabi splitting near 96 \si{\milli \electronvolt} can be clearly identified. The relative vacuum Rabi splitting is estimated to be $\mathrm{16\%}$ of the intersubband transition energy.

\begin{figure}
\centering\includegraphics[width=8cm]{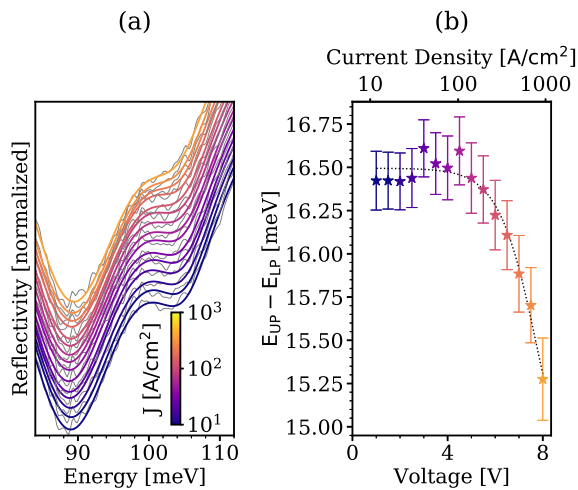}
\caption{The coupling strength is modulated by the current. (a) Spectrally resolved reflectivity of H-resonator array with a single element size of $w=1.60$ \si{\micro \metre}, for different current densities (J). The grey lines are the actual measured spectra and the colorful lines are the Lorentzian curve fits. For high current densities the energy separation between the lower- and upper-polariton absorption peaks is reduced. (b) Separation of the two polariton absorption peaks, extracted from the two Lorentzian curve fits, in dependence of the applied voltage and corresponding current density. The error bars denote one standard deviation of the fitting error, which is significantly smaller than the observed downward trend for current densities beyond 100 \si{\ampere \per \square \centi \metre}.}
\label{fig:couplingdata}
\end{figure}

Applying a current results in a modulation of the intersubband polariton peak positions, such that their energy separation is reduced. Fig. \ref{fig:couplingdata} (a) illustrates the measured spectrally resolved reflectivity of a device array with a single element size of $w=1.60$ \si{\micro \metre} for applied current densities ranging from 10 \si{\ampere \per \square \centi \metre} up to 1000 \si{\ampere \per \square \centi \metre}. The curves are shifted and colored according to the applied current density. The grey curves are the actually measured data and the colored curves are Lorentzian curve fits. It can be seen with the bare eye that for high current densities the separation between the polariton peaks is significantly reduced. Fig. \ref{fig:couplingdata} (b) shows the energy separation between the polariton absorption peaks, extracted from two Lorentzian curve fit peaks. The error bars indicate the single standard deviation fitting errors, which are significantly smaller than the observed variation. It is apparent that for current densities beyond $100$ \si{\ampere \per \centi \metre \squared} the separation between the polariton peaks is reduced rapidly. As the energy separation is a measure for the coupling strength it can be inferred that it is modulated. The larger the minimal energy separation of the polariton absorption peaks, the higher is the coupling strength.

The modulation of the coupling strength can be explained by referring to Eq. \ref{eq:omega1}. The only parameter to allow for electrical modulation is the charge carrier density, when taking into account that the Stark shift is negligible. This underlines that this measurement principle can also be used for optical estimation of the relative charge carrier densities. As a direct consequence of charge conservation, the charging rate of the device is proportional to the difference of the injection and extraction current. The injection is assumed to be independent of the number of charge carriers in the system. Additionally, infinitely large contact reservoirs are assumed. Hence, the injection is completely given by the injection rate $r_\mathrm{in}$. Note that the total extraction is proportional to the number of charge carriers in the system and an extraction rate $r_\mathrm{out}$. This can be written as
\begin{equation}
\label{eq:model}
    \frac{d}{dt}N(t) = \frac{1}{q_0}\left(I_\mathrm{in} - I_\mathrm{out}\right) = r_\mathrm{in}-N(t) r_\mathrm{out},
\end{equation}
where $N(t)$ is the momentary number of charge carriers, $q_0$ is the elementary charge, $I_\mathrm{in}$ is the input current and $I_\mathrm{out}$ is the output current. In the steady-state, which is reached very rapidly due to the low electron transition times, the time derivative of the charge carrier number is zero. Hence the mean number of injected electrons equals the mean number of extracted electrons. As a result the mean number of charge carriers in the system is given by
\begin{equation}
\label{eq:carrier_number}
    N_\mathrm{stat} = \frac{r_\mathrm{in}}{r_\mathrm{out}}.
\end{equation}
A decreasing number of charge carriers, like measured in this experiment, results in faster extraction rates relative to the injection rates. Increasing extraction rates also imply decreasing electron dwell times $\tau_\mathrm{dwell} = r_\mathrm{out}^{-1}$, meaning that the average time an electron spends within the quantum wells is reduced.


We investigate the optical properties of triple barrier resonant tunneling diodes embedded in resonant and off-resonant cavities. A multi-pass measurement is employed to characterize the bare intersubband transitions. Spectrally resolved reflection measurements on arrays of double metal resonators, yield a clear vacuum Rabi splitting, confirming the existence of intersubband polaritons in said devices. Applying electric currents to a device array results in a modulation of the coupling strength, which can be seen in the separation of the lower- and upper-polariton peaks. This underlines that electrical control of strong-coupling can be achieved in devices as simple as resonant tunneling diodes. These results pave the way for electronic devices operating in the strong-coupling regime.

The authors would like to thank Peter Rabl (TU Wien), G\'erald Bastard, Carlo Sirtori, Angela Vasanelli, Yanko Todorov and Mathieu Jeannin (ENS Paris) for stimulating and insightful discussions and acknowledge financial support from the Austrian Science Fund FWF (DK CoQuS W1210 and DK Solids4Fun W1243).

The data that support the findings of this study are available from the corresponding author upon reasonable request.

\nocite{*}
\bibliography{bib}

\end{document}